\documentclass{nature}

\usepackage{amsmath}    
\usepackage{graphicx}   
\usepackage{color}
\usepackage[10pt]{moresize}
\usepackage{graphics,graphicx}
\usepackage{amsfonts}
\usepackage{amssymb}
\usepackage{amscd}
\usepackage{amsmath}
\usepackage{enumerate}
\usepackage{epsfig}
\usepackage{subfigure}

\let\oldcaption\caption
\renewcommand{\caption}{\sffamily \oldcaption}
\begin{document}

\title{Decoy-state quantum key distribution with biased basis choice}

\author{Zhengchao Wei$^{1,2}$, Weilong Wang$^{1,2}$, Zhen Zhang$^{1}$, Ming Gao$^{2}$, Zhi Ma$^{2}$, Xiongfeng Ma$^{1,\ast}$}

\maketitle

\begin{affiliations}
\item
Center for Quantum Information, Institute for Interdisciplinary Information Sciences, Tsinghua University, Beijing, P.~R.~China
\item
State Key Laboratory of Mathematical Engineering and Advanced Computing and Zhengzhou Information Science and Technology Institute, Zhengzhou, Henan, China \\
$^\ast$ To whom correspondence should be addressed. E-mail: xma@tsinghua.edu.cn
\end{affiliations}

\baselineskip24pt

\maketitle

\begin{abstract}
We propose a quantum key distribution scheme that combines a biased basis choice with the decoy-state method. In this scheme, Alice sends \emph{all} signal states in the $Z$ basis and decoy states in the $X$ and $Z$ basis with certain probabilities, and Bob measures received pulses with optimal basis choice. This scheme simplifies the system and reduces the random number consumption. From the simulation result taking into account of statistical fluctuations, we find that in a typical experimental setup, the proposed scheme can increase the key rate by at least $45\%$ comparing to the standard decoy-state scheme. In the postprocessing, we also apply a rigorous method to upper bound the phase error rate of the single-photon components of signal states.
\end{abstract}

\section*{Introduction}
Quantum key distribution (QKD)\cite{Bennett:BB84:1984,Ekert:QKD:1991} is one of the most realistic applications in quantum information. It aims at extending a secret key between two distant parties, commonly noted as Alice and Bob. The unconditional security has been proven even when an eavesdropper, Eve, has unlimited computation power permitted by quantum mechanics\cite{Mayers_01,LoChauQKD_99,ShorPreskill_00,Renner:SecurityPf:2005}.

The best known protocol of QKD is the BB84 protocol\cite{Bennett:BB84:1984} presented by Bennett and Brassard in 1984. In BB84, Alice encodes the key information randomly into the $X$ and $Z$ bases and sends quantum pulses to Bob. Bob measures the received pulses in two bases randomly. After that, they compare the basis through an authenticated classical channel. The key can only be extracted from the pulses where they use the same basis and this results in that on average half of the raw data is discarded. That is, the basis-sift factor is $1/2$ in the original BB84 protocol. This factor can be improved by the efficient BB84 scheme proposed by Lo et al.~\cite{Lo:EffBB84:2005}. In the efficient scheme, Alice and Bob put a bias in the probabilities of choosing the $Z$ basis and $X$ basis, which can make the basis sift-factor close to $100\%$ in the infinitely long key limit. The efficient BB84 scheme is experimentally demonstrated in 2009\cite{Erven:Biased:2009}.

In practical QKD systems, a highly attenuated laser or a weak coherent state source is widely used to substitute for a perfect single-photon source which is beyond state-of-the-art technology. A weak coherent state source contains multi-photon components (details shown in Methods). When multi-photon states are used for QKD, Eve can launch attacks, like the photon-number-splitting (PNS) attack\cite{BLMS:PNS:2000,LutkenhausJahma_02}, to break the security. Since Eve could have a full control of the quantum channel, she can make the transmittance of multi-photon states to be $100\%$ in the PNS attack. In a conventional security analysis\cite{GLLP:2004}, Alice and Bob have to assume all the losses and errors come from the single-photon components in the worst scenario case. As a result, the performance of QKD is very poor. To improve the performance of the weak coherent state QKD, Hwang proposed the decoy-state method\cite{Hwang:Decoy:2003}. Instead of sending one coherent state, Alice sends pulses with different intensities, so that she can obtain more information to monitor the quantum channel. To maintain the detection statistics of coherent states with different intensities, Eve is not able to change the transmittances of single-photon and multi-photon state freely without being noticed by Alice and Bob. The security of the decoy-state method is proven\cite{Lo:Decoy:2005}, along with various practical schemes\cite{Wang:Decoy:2005,MXF:Practical:2005}. Follow-up experimental demonstrations show that the decoy-state method is a very effective way to improve QKD performance\cite{Zhao:DecoyExp:2006,Zhao:Decoy60km:2006,Rosenberg:ExpDecoy:2007,Zeilinger:Decoy:2007,Peng:ExpDecoy:2007}.

Naturally, we can improve the decoy-state method by applying the biased-basis idea of the efficient BB84 protocol. There are a few observations. First, Alice does not need to choose basis when she chooses the vacuum decoy state. Second, if Alice and Bob mainly choose one basis, say $Z$ basis, for key generation, they effectively treat $X$ basis as for quantum channel testing. In this sense, the functionality of $X$ basis is similar to the decoy states. Intuitively, one may expect to combine decoy states and $X$ basis together.

Here, we propose a new decoy-state method with biased basis choice, following the widely used decoy-state scheme, vacuum+weak decoy-state method\cite{MXF:Practical:2005} (a quick review is shown in Methods), where Alice sends out pulses with three different intensities, vacuum (with an intensity of 0), weak decoy (with an intensity of $\nu$) and signal (with an intensity of $\mu$) states.
\begin{enumerate}
\item
Alice prepares \emph{all} the signal pulses ($\mu$) in the $Z$ basis, where the final secure key is extracted from.

\item
She prepares weak decoy pulses ($\nu$) in the $X$ and $Z$ with certain probabilities.

\item
If she chooses the vacuum decoy state, she does not need to set any basis.

\item
Bob measures the received pulses in the $X$ basis and $Z$ basis with probabilities $p_{x}$ and $p_{z}$, respectively.
\end{enumerate}
The scheme is summarized in Table \ref{Tab:Biased:Scheme}. In the new scheme, only 4 sets of preparations are used by Alice. Compared to the original vacuum+weak decoy-state method, where 6 sets are used, the proposed scheme can simplify the system and reduce the cost of random numbers. Later in the simulation, we will show that this scheme can also improve the QKD performance.


Following the GLLP security analysis\cite{GLLP:2004}, the key generation rate\cite{Lo:Decoy:2005,Lo:Vacuum:2005} is given by
\begin{equation} \label{DecoyBiased:Scheme:KeyRate}
\begin{aligned}
R &\geq q\{-I_{ec} + Q_{1}^{z}[1-H(e_{1}^{pz})]+Q_0\}, \\
I_{ec} &= fQ_{\mu}H(E_{\mu}), \\
q &= \frac{N_\mu p_{z}}{N_{total}},
\end{aligned}
\end{equation}
where $q$ is the raw data sift factor, including basis-sift factor and signal-state ratio; $I_{ec}$ is the cost of error correction and the rest terms in the bracket is the rate of privacy amplification; $f$ is the error correction inefficiency; $Q_\mu$ and $E_\mu$ are the overall gain and quantum bit error rate (QBER); $Q_{1}^{z}$ is the gain of the single-photon components and $e_{1}^{pz}$ is its corresponding phase error rate; $Q_0$ is the background gain; $H(x)=-x\log_{2}(x)-(1-x)\log_{2}(1-x)$ is the binary Shannon entropy function. Note that in our scheme, the final key is extracted from $Z$-basis measurement results, so all the variables in equation~\eqref{DecoyBiased:Scheme:KeyRate} should be measured in the $Z$ basis. The phase error rate $e_{1}^{pz}$ cannot be measured directly, which, instead, can be inferred from the error rate in the $X$ basis\cite{MXF:Finite:2011}.

The gain and QBER, $Q_{\mu}$ and $E_{\mu}$, can be measured from the experiment directly. Alice and Bob need to estimate $Q_{1}^{z}$ and $e_{1}^{pz}$ for privacy amplification. According to the model reviewed in Methods, we have $Q_{1}^{z}=Y_{1}^{z}\mu e^{-\mu}$ and $Q_{0}=Y_{0} e^{-\mu}$, where $Y_{1}^{z}$ and $Y_{0}$ are the yield of single-photon components measured in the $Z$ basis and background rate, respectively. In order to lower bound the key rate equation~\eqref{DecoyBiased:Scheme:KeyRate}, one can lower bound $Y_{1}^{z}$, $Y_0$ and upper bound $e_{1}^{pz}$.

Since that both the vacuum state and the single-photon state are basis independent, the yields of vacuum states and single-photon states in different bases are equal
\begin{equation} \label{DecoyBiased:Scheme:Y01xz}
\begin{aligned}
Y_{0}^{x} &=Y_{0}^{z}, \\
Y_{1}^{x} &=Y_{1}^{z}.
\end{aligned}
\end{equation}
While a multi-photon state is basis dependent, whose basis information may be revealed to Eve by, for example, PNS attack\cite{BLMS:PNS:2000}, so for any $i$-photon state ($i\geq2$), in general,
\begin{equation} \label{DecoyBiased:Scheme:Yixz}
\begin{aligned}
Y_{i}^{x}\neq Y_{i}^{z}.
\end{aligned}
\end{equation}
That is, depending on the basis information, Eve may set the yield of $i$-photon state different for the $X$ and $Z$ bases. As for the error rates, the phase error probability in the $Z$ basis equals to the bit error probability in the $X$ basis
\begin{equation} \label{DecoyBiased:Scheme:epzbx}
\begin{aligned}
e_{1}^{pz}=e_{1}^{bx}.
\end{aligned}
\end{equation}
Then, in the finite-key-size situation where statistical fluctuations should be taken into account\cite{MXF:Finite:2011,Fung:Finite:2010}, we have
\begin{equation} \label{DecoyBiased:Scheme:epzbx}
\begin{aligned}
e_{1}^{pz}\approx e_{1}^{bx}.
\end{aligned}
\end{equation}
Given $e_{1}^{bx}$, we can upper bound $e_{1}^{pz}$ by the random sampling argument (details shown in Methods). We need to point out that even though the single-photon state is basis independent, the error rates in two basis may not be the same
\begin{equation} \label{DecoyBiased:Scheme:e1xz}
\begin{aligned}
e_{1}^{x}\neq e_{1}^{z}.
\end{aligned}
\end{equation}
This can be easily seen by considering a simple intercept-and-resend attack where Eve measures all the pulses in the $Z$ basis, and then she will not introduce any additional error in the Z basis $e_{1}^{z}=0$, but $e_{1}^{x}=1/2$.

\section*{Results}
\begin{figure}
\centering
\resizebox{12cm}{!}{\includegraphics{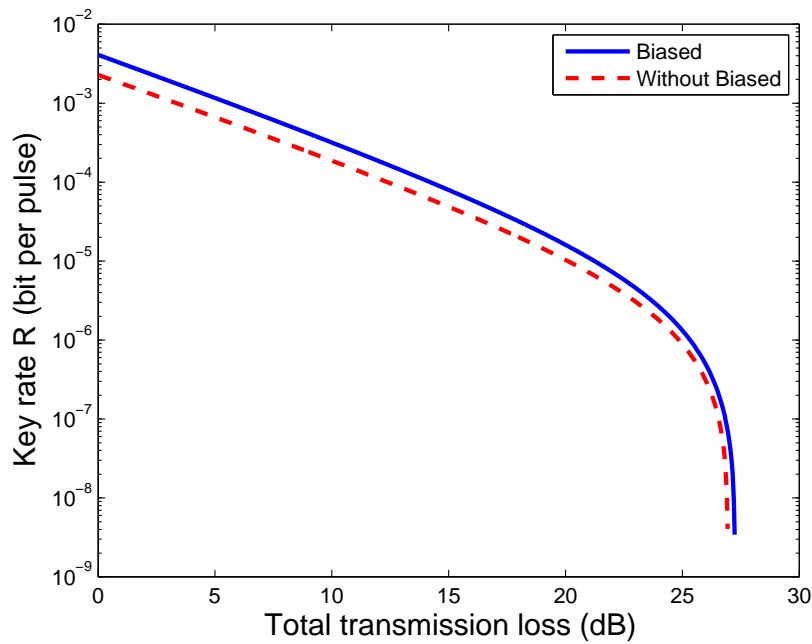}}
\caption{Plot of key rate versus total transmittance. The solid line shows the result of our scheme and the red line shows the result of the standard BB84 with the vacuum+weak decoy-state method. The simulation parameters are shown in Table \ref{Tab:MIFluc:SetupPara}. The confidence interval for statistical fluctuation is 5 standard deviations (i.e., $1-5.73\times 10^{-7}$). The expected photon number of signal state $\mu$ is 0.479. For each transmission loss, we optimize all the parameters, $\nu$, $N_{\mu}$, $N_{\nu}^{z}$,  $N_{\nu}^{x}$, $N_{0}$, $p_z$, and $p_x$.}
\label{Fig:MIFluc:KeyRate}
\end{figure}

In our simulation, the parameters of the experimental setup are listed in Table \ref{Tab:MIFluc:SetupPara}.
Statistical fluctuations are taken into account in the simulation (details shown in Methods). We compare the key generation rate in our scheme with that in the standard BB84 protocol with the vacuum+weak decoy-state scheme. The result is shown in Fig.~\ref{Fig:MIFluc:KeyRate}.

As one can see from Fig.~\ref{Fig:MIFluc:KeyRate}, the key rate of the proposed biased scheme is larger than that of the standard BB84 with vacuum+weak decoy states by at least $45\%$. When the transmission loss is $0$, the key rate improvement can go up to $80\%$. As the transmission loss increases, the improvement of the biased scheme decreases. This is because at a larger transmission loss, more pulses for decoy states are needed and Bob also needs a larger $p_{x}$ to estimate the privacy amplification part in Eq.~\eqref{DecoyBiased:Scheme:KeyRate}. The improvement comes from the fact that $p_x$ is less than $1/2$. As $p_x$ approaches to $1/2$, the biased scheme becomes similar to the original scheme, where $p_x=p_z=1/2$. It is an interesting prospective question how to apply our scheme to QKD systems with high channel losses\cite{PhysRevA.84.062326}.

\begin{figure}
\centering
\resizebox{12cm}{!}{\includegraphics{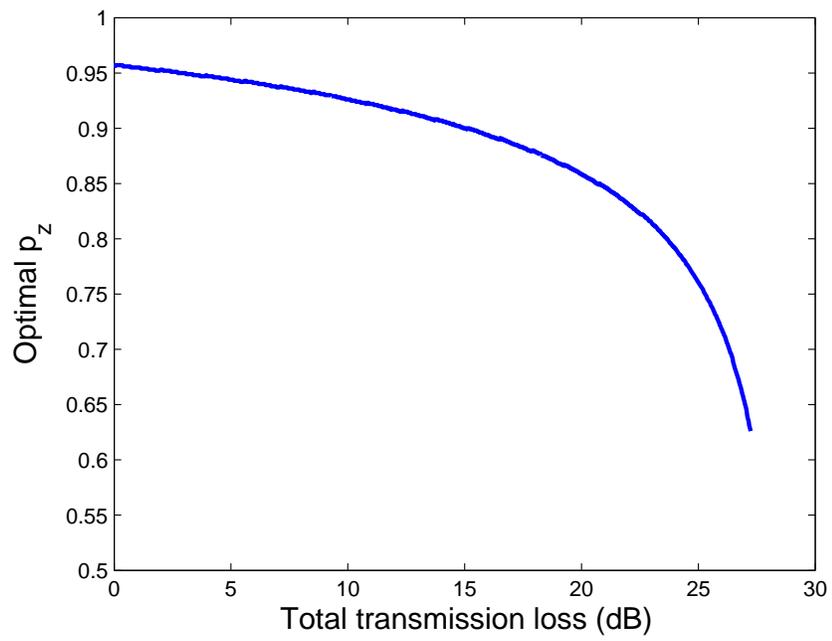}}
\caption{Plot of optimal $p_z$ versus transmission loss. }
\label{Fig:MIFluc:pz}
\end{figure}

For a practical QKD system, one needs to optimize the bias $p_x$ for the key rate. The dependence of the optimal bias on the transmission loss is shown in Fig.~\ref{Fig:MIFluc:pz},
from which we can see that the optimal $p_z$ is about 0.95 when the transmission loss is below $3$ dB and  decreases as the transmission loss increases. The minimal optimal $p_z$ is about 0.6, which is close to 1/2. That is why our scheme approaches the standard BB84 with the vacuum+weak decoy-state scheme as the transmission loss increases.

\section*{Discussion}
In conclusion, we combine the decoy-state QKD with a biased basis choice to enhance the system performance. The key point of our scheme is increasing the raw data sift factor by setting all signal states in one ($Z$) basis. We take statistical fluctuations into account and use a rigorous method to upper bound the phase error rate of the single-photon components of the signal state. Comparing the result with that in the standard decoy-state BB84 protocol, we find an improvement in the key generation rate. Meanwhile, we reduce the complexity of the QKD system by assigning all signal states in the $Z$ basis.

\section*{Methods}

\subsection{Model} \label{App:Biased:Model}
The weak coherent state source is equivalent to a photon-number channel model and its photon number follows a Poisson distribution~\cite{MXF:Practical:2005}:
\begin{equation}\label{DecoyBiased:Appendix:Model:poisson}
\begin{aligned}
P\left( n \right) = \frac{{{\mu ^n}}}{{n!}}{e^{ - \mu }}.
\end{aligned}
 \end{equation}


Define $Y_{i}$ as the yield of an $i$-photon state; $\eta$ as the transmittance of the channel measured in $dB$; $Y_{0}$ as the background count rate. Then, in a normal channel when there is no Eve's intervention, $Y_{i}$ is given by:
\begin{equation}\label{DecoyBiased:Appendix:Model:Yi}
\begin{aligned}
{Y_i} = 1-(1-Y_{0})(1-\eta)^{i}.
 \end{aligned}
 \end{equation}

The gain of $i$-photon states $Q_{i}$ is given by:
\begin{equation}\label{DecoyBiased:Appendix:Model:Qi}
\begin{aligned}
{Q_i} = {Y_i}\frac{{{\mu ^i}}}{{i!}}{e^{ - \mu }}.
 \end{aligned}
 \end{equation}
The overall gain which means the probability for Bob to obtain a detection event in one pulse with intensity $\mu$ is :
\begin{equation}\label{DecoyBiased:Appendix:Model:Qmu}
\begin{aligned}
{Q_\mu } = \sum\limits_{i = 0}^\infty {Q_{i}}= \sum\limits_{i = 0}^\infty  {{Y_i}\frac{{{\mu
^i}}}{{i!}}{e^{ - \mu }}}.
 \end{aligned}
 \end{equation}

The error rate of $i$-photon states $e_{i}$ is given by
\begin{equation}\label{DecoyBiased:Appendix:Model:eiYi}
\begin{aligned}
{e_i}{Y_i} = {{e_0}{Y_0} + {e_d}[1-(1-{\eta})^i](1-Y_{0})},
  \end{aligned}
  \end{equation}
where $e_{d}$ is the probability that a photon hits the erroneous detector and $e_0=1/2$. The overall QBER is given by
\begin{equation}\label{DecoyBiased:Appendix:Model:Emu}
\begin{aligned}
{E_\mu } {Q_\mu }= \sum\limits_{i = 0}^\infty {{e_i}{Y_i}\frac{{{\mu
^i}}}{{i!}}{e^{ - \mu }}} .
 \end{aligned}
 \end{equation}

 Without Eve changing $Y_{i}$ and $e_{i}$, the gain and QBER are given by
 \begin{equation}\label{DecoyBiased:Appendix:Model:QmuEmu}
\begin{aligned}
 {Q_\mu } &= 1 - {e^{ - \eta \mu }}(1-{Y_0}), \\
 {E_\mu }{Q_\mu } &= {e_0}{Y_0} + {e_d}(1 - {e^{ - \eta \mu }})(1-{Y_0}) \\
 \end{aligned}
 \end{equation}

\subsection{Upper bound of $e_{1}^{pz}$} \label{App:Biased:Upe1}
Here, we review the random sampling argument~\cite{Fung:Finite:2010}: using the bit error rate measured in the $X$ basis, $e_{1}^{bx}$, to estimate the phase error rate in the $Z$ basis, $e_{1}^{pz}$, for privacy amplification.

If the key size is infinite, we know that $e_{1}^{bx}=e_{1}^{pz}$. Otherwise, given $e_{1}^{bx}$, $n_{x}$ (the number of decoy states that Alice sends and Bob measures in the $X$ basis), and $n_{z}$ (the number of signal states that Alice sends and Bob measures in the $Z$ basis), we can give a probabilistic upper bound of $e_{1}^{pz}$ such that it is lower than $e_{1}^{pz}$ with a small probability $P_{\theta x}$
\begin{equation} \label{DecoyBiased:Scheme:KeyRate:Pthetax}
\begin{aligned}
P_{\theta x}\equiv Pr\{e_{pz}\geq e_{bx}+\theta_{x}\},
\end{aligned}
\end{equation}
where $\theta_{x}$ is the deviation of the phase error rate from the tested value. Here, $P_{\theta x}$ is a controllable variable and is equal to $10^{-7}$ in the simulation. We have
 \begin{equation} \label{DecoyBiased:Scheme:KeyRate:xi}
\begin{aligned}
P_{\theta x}< \frac{\sqrt{n_{x}+n_{z}}}{\sqrt{e_{bx}(1-e_{bx})n_{x}n_{z}}}2^{-(n_{x}+n_{z})\xi_{x}(\theta_{x})},
\end{aligned}
\end{equation}
where the function $\xi_{x}(\theta_{x})$ is given by
 \begin{equation} \label{DecoyBiased:Scheme:KeyRate:thetax}
\begin{aligned}
\xi_{x}(\theta_{x})\equiv H(e_{bx}+\theta_{x}-q_{x}\theta_{x})-q_{x}H(e_{bx})-(1-q_{x})H(e_{bx}+\theta_{x}),
\end{aligned}
\end{equation}
and $q_{x}=n_{x}/(n_{x}+n_{z})$. Given $P_{\theta x}$, we compute the value of $\xi_{x}$ and  find the value $\theta_{x} $ which is the root of equation~\eqref{DecoyBiased:Scheme:KeyRate:thetax}. We get the probabilistic upper bound \\
 \begin{equation} \label{DecoyBiased:Scheme:KeyRate:e1pzU}
\begin{aligned}
 e_{1}^{pzU}=e_{bx}+\theta_{x}.\\
\end{aligned}
\end{equation}

\subsection{Vacuum+weak decoy state} \label{App:Biased:VWdecoy}
 In this protocol, Alice and Bob use two decoy states to estimate the low bound of $Y_{1}$ and the upper bound of $e_{1}$. First, they implement a vacuum decoy state to estimate the background counts in signal states
\begin{equation}\label{DecoyBiased:Appendix:Vacuum+weak:QvacuumEvacuum}
\begin{aligned}
 {Q_{vacuum}} &= {Y_0}, \\
 {E_{vacuum}} &= {e_0} = \frac{1}{2}. \\
 \end{aligned}
 \end{equation}

Secondly, they perform a weak decoy state where Alice uses a weaker intensity $\nu$ ($\nu<\mu$) for the decoy state to estimate $Y_{1}$ and $e_{1}$. We have:
\begin{equation}\label{DecoyBiased:Appendix:Vacuum+weak:Y1l}
\begin{aligned}
{Y_1} \ge Y^{L}_1 =  \frac{\mu }{{\mu \nu  - {\nu ^2}}}\left(
{{Q_\nu }{e^\nu } - {Q_\mu }{e^\mu }\frac{{{\nu ^2}}}{{{\mu ^2}}} -
\frac{{{\mu ^2} - {\nu ^2}}}{{{\mu ^2}}}{Y_0}} \right),
 \end{aligned}
 \end{equation}
and
\begin{equation}\label{DecoyBiased:Appendix:Vacuum+weak:e1u}
\begin{aligned}
{e_1} \leq e^{U}_1 = \frac{{{E_\nu }{Q_\nu }{e^\nu } -
{e_0}{Y_0}}}{{Y^{L}_1\nu }}.
\end{aligned}
\end{equation}

Note that in our scheme all the parameters for estimating $Y^{L}_1$ are measured in the $Z$ basis and the parameters for estimating $e^{U}_1$ are measured in the $X$ basis. And we must lower bound $Y_{0}$ to obtain the lower bound of the key rate\cite{MXF:Practical:2005}. The $e^{U}_1$ we get here will substitute $e_{bx}$ in equation~\eqref{DecoyBiased:Scheme:KeyRate:Pthetax}.

\subsection{Statistical fluctuation} \label{App:Biased:Fluctuation}
Here, we consider statistical fluctuations for the decoy-sate method~\cite{MXF:Practical:2005}. We need to modify the estimation of $Y_{1}$, equation~\eqref{DecoyBiased:Appendix:Vacuum+weak:Y1l}, and $e_{1}$, equation~\eqref{DecoyBiased:Appendix:Vacuum+weak:e1u}.

The total number of pulses sent by Alice is composed of four cases
\begin{equation}\label{DecoyBiased:Appendix:fluctuation:Ntotal}
\begin{aligned}
N_{total}=N_{\mu}+N_{\nu}^{z}+N_{\nu}^{x}+N_{0}.
 \end{aligned}
 \end{equation}
Since that Alice sends all signal states in the $Z$ basis and the final key is only extracted from the data measured in the $Z$ basis, the parameter $q$ in equation~\eqref{DecoyBiased:Scheme:KeyRate} is given by
\begin{equation}\label{DecoyBiased:Appendix:fluctuation:q}
\begin{aligned}
q=\frac{N_{\mu}p_{z}}{N_{total}}.
 \end{aligned}
\end{equation}
We follow the statistical fluctuation analysis proposed by Ma \emph{et al.}\cite{MXF:MIFluc:2012}.
\begin{equation}\label{DecoyBiased:Appendix:fluctuation:QmuuQnulY0lQ0l}
\begin{aligned}
\begin{array}{l}
 Q_\mu ^U =\hat{Q}_{\mu } (1 + \frac{{u_\alpha }}{\sqrt {{N_\mu }{p_z}{Q_\mu }}}), \\
 Q_\nu ^L = \hat{Q}_{\nu }(1 - \frac{{u_\alpha }}{\sqrt {N_{\nu}^{z}{p_z}{Q_\nu }}}), \\
 Y_0^L = \hat{Y}{_0}(1-\frac{{u_\alpha }}{\sqrt {{N_0}{Y_0}}}), \\
 Q_{0}^{L}=Y_{0}^{L} e^{-\mu}(1-\frac{{u_\alpha }}{\sqrt {{N_0}{Q_0}}}),
 \end{array}
 \end{aligned}
 \end{equation}
where $\hat{Q}_{\mu }$, $\hat{Q}_{\nu }$ and $\hat{Y}{_0}$ are measurement outcomes which means that they are rates instead of probabilities. If we follow the standard error analysis assumption, $u_{\alpha}$ is the number of standard deviations one chooses for the statistical fluctuation analysis. Note that $Q_\mu ^U$ and $Q_\nu ^L$ are used to estimate $Y^{L}_1$, so they should be measured in the $Z$ basis. Here we use equation~\eqref{DecoyBiased:Scheme:KeyRate:Pthetax} to estimate the upper bound of $e_{1}^{pz}$ with
\begin{equation}\label{DecoyBiased:Appendix:fluctuation:nxnz}
\begin{aligned}
n_{z}=N_{\mu}p_{z}Y_1^L\mu e^{-\mu},\\
n_{x}=N_{\nu}^{x}p_{x}Y_1^L\nu e^{-\nu}.
 \end{aligned}
 \end{equation}

\noindent\textbf{Acknowledgments}\\
This work is supported by National Basic Research Program of China Grants No.~2011CBA00300 and No.~2011CBA00301, National Natural Science Foundation of China Grants No.~61073174, No.~61033001, No.~61061130540 and No.~U1204602, the 1000 Youth Fellowship program in China, and National High-Tech Program of China Grant No.~2011AA010803.

\noindent\textbf{Author Contributions}\\
ZW, WW, ZZ, MG, ZM and XM all contributed equally to this paper.

\noindent\textbf{Additional Information}\\
The authors declare no competing financial interests.

\bibliographystyle{apsrev4-1}


\begin{table}
\centering
\caption{List of Alice and Bob's operations. Alice prepares and sends $N_{total}$ pulses, with $N_{total}=N_{\mu}+N_{\nu}^{z}+N_{\nu}^{x}+N_{0}$. Bob measures the received pulses with certain probabilities, $p_{z}+p_{x}=1$.} \label{Tab:Biased:Scheme}
\begin{tabular}{ccc|ccc}
\hline\hline
Alice prepares and sends &&& Bob measures\\
\hline
$N_{\mu}$ signal pulses in the $Z$ basis&&& \\
$N_{\nu}^{z}$ decoy pulses in the $Z$ basis&&& \quad with probability $p_{z}$ in the $Z$ basis\\
$N_{\nu}^{x}$ decoy pulses in the $X$ basis&&& \quad with probability $p_{x}$ in the $X$ basis\\ $N_{0}$ vacuum pulses&&& \\
\hline\hline
\end{tabular}
\end{table}

\begin{table}
\centering
\caption{List of experimental parameters for simulation.} \label{Tab:MIFluc:SetupPara}
\begin{tabular}{cccccccccccc}
\hline\hline
$N_{total}$ &&& $f$ &&& $e_d$ &&&$Y_0$ \\
\hline
$6\times10^{9}$ &&& $1.16$ &&& $3.3\%$ &&& $1.7\times10^{-6}$ \\
\hline\hline
\end{tabular}
\end{table}

\clearpage

\clearpage

\end{document}